\documentstyle[11pt]{article}
\catcode`\@=11
\begin{document} \openup6pt

\title{Probability for Primordial Black Holes in Multidimensional Universe with Nonlinear Scalar Curvature Terms }

\author{B. C. Paul$^1$\thanks{E-mail: bcpaul@iucaa.ernet.in }, A. Saha\thanks{E-mail: arindamjal@gmail.com} $^2$ and S. Ghose\thanks{E-mail: souviknbu@rediffmail.com}$^1$\\
$^1$Physics Department, North Bengal University \\
   Darjeeling, PIN : 734 430, India. \\
 $^2$Department of Physics, Darjeeling Govt. College \\
 Darjeeling , Pin : 734 101, India.  }
\date{}

\maketitle

\begin{abstract}
We investigate multi-dimensional universe with nonlinear scalar curvature terms to evaluate the probability of creation of primordial black holes. For this we obtain Euclidean instanton solution in two different topologies: (a) $S^{D-1}$ - topology which does not accommodate primordial black holes and (b) $S^1\times S^{D-2}$-topology which accommodates a pair of black holes. The probability for quantum creation of an inflationary universe with a pair of black holes has been evaluated assuming a gravitational action which is described by a polynomial function of scalar curvature with or without a cosmological constant ($\Lambda $) using the framework of semiclassical approximation of Hartle-Hawking boundary conditions. We discuss here a class of new gravitational instantons solution in the $R^4$-theory which are relevant for cosmological model building. \\

{\it PACS No(s) : 04.70 Dy, 98.80.Bp, 98.80Cq}
\end{abstract}

\pagebreak

{\bf 1. INTRODUCTION :}

  The last couple of decades witnessed a considerable progress in modern cosmology  both theoretically [1] and observationally [2]. It has been generally accepted that inflation is one of the main criteria to build a consistent cosmological models of the early universe. The important outcome  of inflation is that it explains satisfactorily the observed large scale structure of the universe in addition to solving longstanding problems both in particle physics and cosmology.  The quantum fluctuation of the fields in the early universe may produce  abundantly primordial black holes (in short PBH). 
  In the literature the problem of creation and subsequent evolution  of PBH in the early universe are discussed  in different contexts [3]. In 1975, the discovery of Hawking radiation ushered in a new era in black hole physics [4]. The life times of black holes formed by gravitational collapse of a massive star are quite long which are in fact comparable to the age of the universe. However, primordial black holes are tiny and are comparatively short lived. Therefore, the hope of confirming Hawking radiation by observation is through PBH hunting. The existance of a pair of PBH may be important in the context of the recent cosmological observations. Bousso and Hawking (BH) [5] calculated the probability of the quantum creation of a universe with a pair of primordial black holes in a $4$ dimensional universe considering two different Euclidean space-times : (i) a universe with space-like sections with $S^3$- topology and (ii) a universe with space-like section with $S^1 \times S^2$- topology as is obtained in the Schwarschild-de Sitter solution. The first one describes an inflationary universe without a pair of PBH, while the later describes a Nariai universe [6], an inflationary universe with a pair of black holes. BH considered a massive scalar field to investigate the problem which provided an effective cosmological constant for a while through a slow rolling potential (mass term) in the early epoch. However, the evolution of a universe at the Planck time $t\sim M_P^{-1}$ may be better understood in the framework of quantum gravity. A consistent quantum theory of gravity is yet to emerge. It was Kaluza and Klein [7] who first independently initiated the study to formulate a unified theory of gravity and electro-magnetic interaction by introducing an extra dimension. But the KK-approach did not work well subsequently because of uninteresting dimensional reduction technique employed in the theories. In recent years there is a paradigm shift in cosmological model building in the higher dimensions from that of previous approaches. The success of Superstring theory/M-theories revived the studies in higher dimensional framework which has been considerably generalized [8]. In modern higher dimensional scenario, the fields of the standard model  are considered to be confined to (3+1) dimensional hyper-surface (referred to as 3-Brane) embedded in a higher dimensional space-time but the gravitational field may propagate through the bulk dimensions perpendicular to the brane which is referred to as braneworld [9]. At high enough energy scale, Einstein field equation in 4 dimensions may not give a correct predictions of the universe. Such an energy scale perhaps was available shortly after the big bang. Thus a more general theory is needed which at low energy reproduces the Einstein's  theory. In the usual 4-dimensions or multidimensional universe the gravitational sector of the action may be modified by considering non-linear terms of scalar curvature ($R$).  
  The motivation of studying higher dimensions is that most of the theories of particle interactions, including String theory requires space-time dimensions more than the usual four for their consistent formulation.  It is therefore considered essential to check if consistent cosmological or astrophysical solutions, which can accommodate these theories are also allowed. 
  In this paper we consider such a modified theory in multidimensional universe to evaluate the probability of pair creation of PBH using the prescription of Hartle-Hawking no boundary proposal [10]. According to the no boundary proposal, the quantum state of the universe is defined by path integrals over Euclidean metrics $g_{\mu\nu }$ on compact manifolds $M$. From this it follows that the probability of finding a three metric $h_{ij}$ on the spacelike surface $\partial M$ is given by a path integral over all $g_{\mu\nu }$ on $M$ that agrees with $h_{ij}$ on $\partial M$. With the inclusion of matter fields, one obtains the wave function of the universe which is given by
\begin{equation}
\Psi[h_{ij}, \Phi_{\partial M}]= \int D(g_{\mu\nu},\Phi) \; exp[-I(g_{\mu\nu},\Phi)],
\end{equation}
where $(h_{ij}, \Phi _{\partial M})$ are the three-metric and matter fields on a spacelike boundary $\partial M$ and the path integral is taken over all compact Euclidean metrics $g_{\mu\nu}$. 
We now estimate $\Psi$ from a saddle point approximation to the path integral. Two types of topologies that describe inflationary universe characterized by spacelike sections, namely (i) $R\times S^d$ -topology and (ii) $R\times S^1\times S^d$ -topology are considered. The former represents a universe without a pair of PBH and the later represents that with a pair of PBH. We obtain Euclidean solutions of the modified Einstein equation corresponding to each of these universes, which can be analytically continued to match a boundary $\partial M$ of the appropriate topology.  The Euclidean action $I$ is evaluated for the above solutions. The wave function of the universe in the semiclassical approximation is given by 
\begin{equation}
\Psi[h_{ij}, \Phi_{\partial M}]\approx \sum _{n} A_{n}e^{-I_{n}},
\end{equation}
where the sum is over the saddle points of the path integral, and $I_{n}$ denotes the corresponding Euclidean action. We can thus assign a probability measure to each type of universe:
\begin{equation}
P[h_{ij}, \Phi_{\partial M}] \sim e^{\left(-2I^{Re}\right)},
\end{equation}
where the superscript $Re$ denotes the real part of the action corresponding to the dominant saddle point, i.e., the classical solution satisfying the Hartle-Hawking (HH) boundary conditions [10].\\ 
It is well known that a theory with higher order Lagrangian is conformally equivalent to Einstein gravity with a matter sector containing a minimally coupled, self interacting scalar field. However, the renormalisation of higher loop contributions introduces terms into the effective action that are higher than quadratic in R. Consequently it is important to study the effects of these terms in the quantum creation of a universe with a pair of PBH. Paul {\it et al.} [11]   studied the probability of pair creation including $R^2$ term in  four dimensions. Later  an effective action upto $R^3$ term was considered in {\it Ref.} [12] and the probability of creation of a pair of black holes in the modified Einstein action in $4$-dimensions is estimated.  It is noted that  the probabilty of creation of a universe with a pair of PBH suppresses in the $R^3$-theory without a cosmological constant. Subsequently, a higher dimensional universe was considered in Einstein-Hilbert action to probe the PBH pair creation rate [13]. G\"unther {\it et al.} [14] studied multidimensional gravitational models with scalar curvature non-linearities of types $R^{-1}$ and $R^4$ to study the existance of at least one minimum of the effective potential for the volume moduli of the internal spaces with an emphasis to obtain stabilization. The stabilization of the internal space is important because the extra dimesional space components should be static or nearly static at least from the time of primordial nucleosynthesis. In contrast to $R^2$-models, the $R^4$-model shows a rich substrate of the stability region in parameter space which minimally depends on the total dimension of the bulk space. In this paper we consider a modified gravitational action described by a Lagrangian polynomial in $R$  considering terms up to  $R^4$ 
to probe the early universe. 

 The paper is organised as follows : in section 2 we consider  the action and  the correspoding field equation is determined. We then obtain gravitational instanton solutions in two different topologies. In section 3 we use the action to estimate the relative probability of the two types of universes with different topologies in the presence of PBH. In section 4 we give a brief discussion.

\pagebreak

{\bf 2. Gravitational Instantons with or without PBH :}\\

 We consider a higher dimensional Euclidean gravitational action which is given by
\begin{equation}
I_{E} =- \frac{1}{16\pi} \int{ d^{D}x \sqrt{g} f(R) -\frac{1}{8\pi} \int _{\partial M} d^{(D-1)}x \sqrt{h} Kf'(R)},
\end{equation}
where $g$ is the D-dimensional Euclidean metric, $R$ is the Ricciscalar, $\Lambda$ is the cosmological constant, and $K=h^{ab}K_{ab}$ is the trace of the second fundamental form of the boundary $\partial M$ in the metric. We consider a polynomial function in Ricci scalar $f(R)=\sum_{i}\lambda_{i}R^i$, in particular $\lambda_{o}=-2\Lambda, \lambda_{1}=1, \lambda_{2}=\alpha, \lambda_{3}=\beta, \lambda_{4}=\gamma$, i.e., $f(R)=R+\alpha R^2+\beta R^3+\gamma R^4-2\Lambda$.

{\bf A. $S^{D - 1}$- Topology, the de Sitter spacetimes:}\\
In this section, we study vacuum solutions of the field equation corresponding to the non-linear higher dimensional action (4). We look for a solution with space like section $S^{D - 1}$. We choose the D-dimensional metric ansatz, which is given by
\begin{equation}
dS^2=d{\tau}^2+a^2(\tau)d \Omega_{d}^2,
\end{equation}
where $d=D-1$, $a$ is the scale factor of a D dimensional universe and $d \Omega_{d}^2$ is a line element of unit $(D - 1)$-sphere. The scalar curvature is given by 
\begin{equation}
R=-\left[2d\frac{\ddot a}{a}+d(d-1)(\frac{{\dot a}^2}{a^2}-\frac{1}{a^2})\right],
\end{equation}
where the overdots denotes differentiation with respect to $\tau $.
We use the constraint through a Lagrangian multiplier $ \tilde{\beta} $ and rewrite the action as
\begin{equation}
I_{E}=-V_{o}\int\left[f(R)a^d-\tilde{\beta}\left(R+2d\frac{\ddot a}{a}+d(d-1)\left(\frac{{\dot a}^2}{a^2}-\frac{1}{a^2}\right)\right)\right]d\tau-\frac{1}{8\pi}\int_{\delta M}{d^{(D-1)}x\sqrt{h} f'(R)},
\end{equation}
where $V_{o}=\frac{1}{16\pi} \frac{2\pi^{(d+1)/2}}{\Gamma(d+1)/2}$.
Variaton of the action with respect to $R$, determines $\tilde{\beta}$, which is 
\begin {equation}
\tilde{\beta} = a^d f'(R).
\end{equation}
Substituting the above constraint in the action we get
\begin{eqnarray}
I_{E}=-V_{o}\int_{\tau=0}^{\tau_{\delta M}}a^d\left[f(R)-f'(R)\left(R-d(d-1)\left(\frac{\dot a^2}{a^2}+\frac{1}{a^2}\right)\right)+2d  \dot R f''(R)\frac{\dot a}{a} \right] d\tau +{}\nonumber\\  {} \left[2d V_{o} \dot a a^{d-1} f'(R)\right]^{\tau_{\delta M}}_{\tau=0},
\end{eqnarray}
Varying the above action with $a$ and  $R$ respectively, we get  
\begin{eqnarray}
f'(R)\left[2(d-1)\frac{\ddot a}{a} + (d-1)(d-2)\left(\frac{{\dot a}^2}{a^2}-\frac{1}{a^2}\right)+R \right]{}\nonumber\\  {}+f''(R)\left[(d-1)\frac{\dot a}{a}\dot R +2\ddot  R\right]+ 2f'''(R) \dot R^2 - f(R)  =  0,
\end{eqnarray}
\begin{equation}
f''(R)\left[2d\frac{\ddot a}{a}+d(d-1)\left(\frac{{\dot a}^2}{a^2}-\frac{1}{a^2}\right)+R\right]=0.
\end{equation}
The above field equations admits an instanton solution given by 
\begin{equation}
a=\frac{1}{H_{o}}\;\sin{(H_{o}\tau)},
\end{equation}
where $H_{o}$ is determined from the constraint equation 
\begin{equation}
d(d-1)H_{o}^2+ \alpha d^2(d+1)(d-3)H_{o}^4+\beta d^3(d+1)^2(d-5)H_{o}^6+\gamma d^4(d+1)^3(d-7)H_{o}^8-2\Lambda =0.    
\end{equation}
Thus $H_{o}=f(d,\alpha,\beta,\gamma,\Lambda)$ determined by five parameters. We note that for (i) $D=4, \; H_{o}=f(d,\beta,\gamma,\Lambda)$; (ii) $D=6, \; H_{o}=f(d,\alpha,\gamma,\Lambda)$; and (iii) $D=8, \; H_{o}=f(d,\alpha,\beta,\Lambda)$. The above instanton solutions satisfy the HH boundary condition viz., $a(0)=0$, $\dot a(0)=0$. Now one can choose a path along the $\tau ^{\rm Re}$ axis to $\tau = \frac{\pi}{2H}$, the solutions describe half of the Euclidean de Sitter instanton in  $S^{D - 1}$ -topology. Analytic continuation of the metic (5) to Lorentzian region ${x_{1} \to \frac{\pi}{2} + i \sigma }$ gives 
\begin{equation}
ds^2 = d\tau ^2 + a^2(\tau )\left[-d\sigma ^2 + {cosh }^2\sigma \;d\Omega _{d-2}^2   \right],
\end{equation}
which is a de Sitter like metric. However, if one sets $\tau = it$ and $\sigma = \frac{i\pi}{2} + \chi$, the metric becomes 
\begin{equation}
ds^2 = -dt^2 + S^2(t)\left[d\chi ^2 + {{sinh }^2\chi } \;d\Omega _{d-2}^2 \right],
\end{equation}
where $S(t) = -ia(it)$. The above metric now describes an open universe. Thus, the creation of an open inflationary universe may be realized in this case. It may be pointed out here that a closed universe created quantum mechanically may be realized as an open inflationary universe under a Wick rotation. Since it is not yet known whether our universe is exactly flat, it may be interesting to study an open universe. The real part of the Euclidean action corresponding to the solution calculated by following the complex contour of $\tau$ suggested by BH is considered here which determines 
\begin{equation}
I^{Re}_{E}=-\frac{V_{o}I_{d}}{H_{o}^{d+1}}\left[d(d+1)H_{o}^2+\alpha d^2(d+1)^2H_{o}^4+\beta d^3(d+1)^3H_{o}^6+\gamma d^4(d+1)^4H_{o}^8-2\Lambda \right],
\end{equation}
where $I_{d}=\int_{0}^{\frac{\pi}{2H_{o}}}{\sin^d y  \; dy}$, denoting $y=H_{o}\tau$. The value of the integral $I_{d}$ depends on the dimensions of the universe. For odd number of dimensions (D),
\begin{equation}
I_{d}= \frac{d-1}{d}\frac{d-3}{d-2} ........ \frac{3}{4}\frac{1}{2}\frac{\pi}{2},\;\;for\;\;d\;\;even,
\end{equation}
and for even number of dimensions (D),
\begin{equation}
I_{d}= \frac{d-1}{d}\frac{d-3}{d-2} ....... \frac{4}{5}\frac{2}{3}1,\;\;for\;\;d\;\;odd.
\end{equation}
With the choosen path for $\tau $, the solution describes half the de Sitter instanton in a higher dimensional universe with $S^d$ topology, joined to a real Lorentzian hyperboloid of $(R^1\times S^d)$ -topology. It can be joined to any boundary satisfying the condition $a_{\partial M}>0$. For $a_{\partial M}>H_{o}^{-1}$, the wave function oscillates and predicts a classical space-time.

{\bf B.  $S^1 X S^{D - 2}$- Topology :}

We consider vacuum solution of the field equation corresponding to the action (4) and look for a universe with $S^1 X S^{D - 2}$ -spacelike sections. The above topology accomodates a pair of black holes. The corresponding metric ansatz for $D=1+1+d$ dimensions is given by
\begin{equation}
ds^2=d\tau^2+ a^2(\tau)dx^2+b^2(\tau)d\Omega_{d}^2,
\end{equation}
where $a(\tau)$ is the scale factor of two sphere and $b(\tau)$ is the scalefactor of the $^{D - 2}$-sphere given by the metric $$d\Omega_{d}^2=dx_{1}^2+sin^2x_{1}dx_{2}^2+sin^2x_{1}sin^2x_{2}dx_{3}^2+....\;d\; space.$$
The scalar curvature is given by
\begin{equation}
R=-\left[2\frac{\ddot a}{a}+2d\frac{\ddot b}{b}+d(d-1)\left(\frac{{\dot b}^2}{b^2}-\frac{1}{b^2}\right)+2d\frac{\dot a\dot b}{ab}\right].
\end{equation}
Using the above constraint (20) the Eucledian action (4) may be rewritten as
\begin{eqnarray}
I_{E} & = & -V_{o}'\int{\left[f(R)ab^d-\tilde{\beta}\left(R+2\frac{\ddot a}{a}+2d\frac{\ddot b}{b}+d(d-1)\left(\frac{{\dot b}^2}{b^2}-\frac{1}{b^2}\right)+2d\frac{\dot a\dot b}{ab}\right)\right]d\tau}- {}\nonumber\\&& {}\frac{1}{8\pi}\int_{\delta M}{d^{(D-1)}x\sqrt{h} f'(R)},
\end{eqnarray}
where 
$V_{o}'=\frac{1}{16\pi} \frac{2\pi^{(D+1)/2}}{\Gamma \left((D+1)/2\right)}$. The undetermined multiplier $\tilde{\beta}$ may now be obtained by varying the action and inserting the corresponding $\tilde{\beta}$ in the action, we finally obtain
\begin{eqnarray}
I_{E}=-V_{o}'\int_{\tau=0}^{\tau_{\delta M}}ab^d\left(f(R)-f'(R)\left(R-d(d-1)\frac{{\dot b}^2}{b^2}-2d\frac{\dot a\dot b}{ab}-d(d-1)\frac{1}{b^2}\right)\right)d\tau -{}\nonumber\\ {}V_{o}'\int_{\tau=0}^{\tau_{\delta M}}2\dot R f''(R)\left(\frac{\dot a}{a}+ d\frac{\dot b}{b}\right)d\tau + 2ab^df'(R)V_{o}'\left[\frac{\dot a}{a}+d\frac{\dot b}{b}\right]_{\tau=0}^{\tau_{\delta M}}.
\end{eqnarray}
Varying the  action with respect to   $a$,  $b$ and $R$ respectively, we get the following field equations :
\begin{equation}
f''(R)\left[R+2\frac{\ddot a}{a}+2d\frac{\ddot b}{b}+2d\frac{\dot a\dot b}{ab}+d(d-1)\left(\frac{{\dot b}^2}{b^2}-\frac{1}{b^2}\right)\right]=0,
\end{equation}
\begin{eqnarray}
f'(R)\left[R+2d\frac{\ddot b}{b}+d(d-1)\left(\frac{{\dot b}^2}{b^2}-\frac{1}{b^2}\right)\right]+2f''(R)\left(\ddot R+2d\dot R\frac{\dot b}{b}\right)+{}\nonumber\\ {}2{\dot R}^2f'''(R)-f(R)=0,
\end{eqnarray}
\begin{eqnarray}
f'(R)\left[R+2(d-1)\frac{\ddot b}{b}+2(d-1)\frac{\dot a\dot b}{ab}+(d-1)(d-2)\left(\frac{{\dot b}^2}{b^2}-\frac{1}{b^2}\right)+2\frac{\ddot a}{a}\right]+{}\nonumber\\ {}2f'''(R){\dot R}^2+2f''(R)\left[(d-1)\dot R \frac{\dot b}{b}+\ddot R+\dot R\frac{\dot a}{a}\right]-f(R)=0.
\end{eqnarray}
We now look for instanton solutions from the above field equations which is given by
\begin{equation}
a=\frac{1}{H}\sin(H\tau),\;  b=\sqrt{d-1} \;{H^{-1}},\; R=(2+d)H^2
\end{equation}
where $H$ satisfies the following constraint equation
\begin{equation}
d \; H^2+\alpha(2+d)(d-2)H^4+\beta(2+d)^2(d-4)H^6+\gamma (2+d)^3(d-6)H^8-2\Lambda =0.
\end{equation}
Here $H$ depends on the parameters $\alpha, \beta, \gamma, \Lambda$ i.e., $H=f(d,\alpha,\beta,\gamma,\Lambda)$ for a given extra space dimension (d). We note that for (i) $D=4, \; H=f(d,\beta,\gamma,\Lambda)$; (ii) $D=6, \; H=f(d,\alpha,\gamma,\Lambda)$; and (iii) $D=8, \; H=f(d,\alpha,\beta,\Lambda)$.
The above  solution satisfies the HH boundary conditions $a(0)=0,\; \dot a(0)=0,\; b(0)=0,\;  \dot b(0)=0$. Analytic continuation of metric (19) to Lorentzian region i.e., using Wick rotations ${\tau \to it}$ and ${x \to \frac{\pi }{2} + i \sigma}$,  one gets 
\begin{equation}
ds^2 = -dt^2 + S(t)^2\; d\sigma ^2 + H^{-2}\;d\Omega _{d-2}^2,
\end{equation}
where $S(t)=-ia(it)$. In this case it is evident that the analytic continuation of time and space is not an open universe. The space-time now represents an anisotropic universe.  The corresponding Lorentzian solution is given by \\ 
$a(\tau ^{\rm Im})|_{\tau ^{\rm Re}= \frac{\pi }{2H}} = H^{-1}{cosh H\tau ^{\rm Im}}$ \\
$b(\tau ^{\rm Im})|_{\tau ^{\rm Re} = \frac{\pi }{2H}} = H^{-1}$ \\
Its spacelike section represents $(D-2)$-spheres of radius $a$ with a hole of radius $b(=H_{o}^{-1})$ punched through the north and south poles. The physical interpretation of the solution is that of $(d-1)$ spheres containing two blackholes, they accelerate away from each other with the expansion of the universe, at the opposite ends. The real part of the action may now be determined following the contour approach suggested by BH, which is 
\begin{equation}
I_{S^1\times S^d}^{Re}=-\left[\frac{V_{o}'(d-1)^{d/2}}{H^{d+2}}\left((2+d)H^2+\alpha (2+d)^2H^4+\beta (2+d)^3H^6+\gamma (2+d)^4H^8-2\Lambda \right)\right],
\end{equation}
where $d=D-2$. The solution (26) describes a universe with two black holes at the poles of a $(D-2)$-sphere. It may be pointed out here that the non-zero contribution to the action comes from the surface terms only.

{\bf 3. Evaluation of the probability for PBH : } \\

In the previous section we compute the action for inflationary universe with or without a pair of black holes. The probability for creation of a higher dimensional de Sitter universe in $f(R)$-theory may now be obtained using gravitational instantons obtained in the previous sections. The probability for nucleation of a higher dimensional universe without PBH is 
\begin{equation}
P_{S^{D-1}}\sim e^ {\left[\frac{2V_{o}I_{D-1}}{H_{o}^D}\left[D(D-1)H_{o}^2+\alpha D^2(D-1)^2H_{o}^4+\beta D^3(D-1)^3H_{o}^6+\gamma D^4(D-1)^4H_{o}^8-2\Lambda \right]\right]},
\end{equation}
where $H_{o}$ satifies the constraint eq.(13). However, the probability of nucleation of  an inflationary universe with a pair of black holes is obtained from action $(22)$, which is 
\begin{equation}
P_{S^1\times S^{D-2}}\sim e^ {\left[\frac{2V_{o}'(D-3)^{(D-2)/2}}{H^D}\left[DH^2+\alpha D^2H^4+\beta D^3H^6+\gamma D^4H^8-2\Lambda \right] \right]},
\end{equation}
where $H$ satifies constraint eq.(27).\\

{\bf Special Cases :}
For $D\geq4$ we note the following :\\ 
(i)$\;\alpha=\beta=\gamma =0$, the probability estimates given in (30) and (31) reduce to
\begin{equation}
P_{S^{D-1}}\sim e^ {\left[4V_{o}I_{D-1}(D-1)^{D/2}\left(\frac{D-2}{2\Lambda}\right)^{(D-2)/2}\right]},
\end{equation}
\begin{equation}
P_{S^1\times S^{D-2}}\sim e^{\left[4V_{o}'\left(\frac{(D-3)(D-2)}{2\Lambda}\right)^{(D-2)/2}\right]}.
\end{equation}
We recover the probabilities obtained by Bousso and Hawking in Einstein theory when 
$D=4$, which is 
\begin{equation}
 P_{S^3}\sim e^{3\pi/\Lambda}, \; P_{S^1\times S^2}\sim e^{2\pi/\Lambda}.
\end{equation}
In the above de Sitter universe is more probable for a positive cosmological constant.\\
(ii) For $\alpha=\beta = 0,\; \gamma \neq 0$, the corresponding probabilities are 
\begin{equation}
P_{S^{D-1}}\sim e^ {\left[\frac{2V_{o}I_{D-1}}{H_{o}^D}\left(\frac{-6D(D-1)H_{o}^2+16\Lambda }{D-8}\right)\right]},
\end{equation}
\begin{equation}
P_{S^1\times S^{D-2}}\sim e^{\left[\frac{2V_{o}'(D-3)^{(D-2)/2}}{H^D}\left(\frac{-6DH^2+16\Lambda }{D-8}\right)\right]},
\end{equation}
where $H_{o}$ and $H$ are determined from eqs. (13) and (27) respectively for $D \neq 8$. 
In four dimensions the probabilities reduce to
\begin{equation} 
P_{S^3}\sim e^{\left[\frac{\pi}{3H_{o}^4}(9H_{o}^2-2\Lambda)\right]}, \; \;P_{S^1\times S^2}\sim e^{\left[\frac{2\pi}{H^4}(3H^2-2\Lambda)\right]}.
\end{equation}
For $\Lambda=0$ and $D\neq8$, we evaluate the probabilities which are 
\begin{equation}
P_{S^{D-1}}\sim e^ {\left[12V_{o}I_{D-1}(D(D-1))^{D/2}(8-D)^{(D-8)/6}(\frac{\gamma}{D-2})^{(D-2)/6}\right]},
\end{equation} 
\begin{equation}
P_{S^1\times S^{D-2}}\sim e^{\left[12V_{o}'D^{D/2}(8-D)^{(D-8)/6}(D-3)^{(D-2)/2}(\frac{\gamma}{D-2})^{(D-2)/6}\right]}.
\end{equation} 
We note the following: \\
 (i) a new gravitational instanton solution in higher dimensions is obtained for a $R^4$-theory with a negative coupling parameter  (   i.e., $\gamma <0$) for $D>8$, and that with a positive (i.e., $\gamma >0$ ) for $D<8$, (ii)  de Sitter universe  is more probable for $D\geq 3$.\\  
We also note that for $D=8$, an  instanton solution with $H_{o}^2=\frac{\Lambda}{21}$ in $S^7$-topology and with    $H^2=\frac{\Lambda}{3}$ in  $S^1\times S^6$-topology  are permitted. The probabilties of PBH for $D=8$ dimensions are 
\begin{equation}
P_{S^7}\sim e^{\left[\frac{1372\pi^3}{15\Lambda^3}\left(27+2048\gamma \Lambda^3\right)\right]}, P_{S^1\times S^6}\sim e^{\left[\frac{100\pi^3}{21\Lambda^3}\left(27+2048\gamma \Lambda ^3\right)\right]}.
\end{equation}
We further note that the probabilities are same if  (i)  $|\gamma|= \frac{27}{2048\Lambda^3}$  and $\Lambda > 0$ with negative $ \gamma$ and (ii) $\gamma=\frac{27}{2048\Lambda^3}$ and $\Lambda < 0$ with positive $ \gamma$. 
An interesting  gravitational instanton is permitted without a cosmological constant in $D= 8$.

(iii) For $\alpha \neq 0,\;\beta =\gamma =0,$ the probabilities are 
\begin{equation}
P_{S^{D-1}}\sim e^{\left[\frac{2V_{o}I_{D-1}}{H_{o}^D}\left(\frac{-2D(D-1)H_{o}^2+8\Lambda}{D-4}\right)\right]},
\end{equation}
\begin{equation}
P_{S^1\times S^{D-2}}\sim e^{\left[\frac{2V_{o}'(D-3)^{(D-2)/2}}{H^D}\left(\frac{-2DH^2+8\Lambda }{D-4}\right)\right]}.
\end{equation}
In this case PBH pair creation may be possible even without a cosmological constant. For $\Lambda =0$, the probabilities are given by
\begin{equation}
P_{S^{D-1}}\sim e^ {\left[\frac{4V_{o}I_{D-1}}{(D-2)^{(D-2)/2}}(D(D-1))^{D/2}(4-D)^{(D-4)/2}\right]},
\end{equation}
\begin{equation}
P_{S^1\times S^{D-2}}\sim e^ {\left[4V_{o}'\left(\frac{D-3}{D-2}\right)^{(D-2)/2}D^{D/2}(4-D)^{(D-4)/2}\right]}.
\end{equation}
We note that the instanton solutions considered here for the two different topologies physically sensible when $\alpha <0$ for $D>4$ and $\alpha >0$ for $D<4$. We note that in higher dimensions with space-time dimensions $D\geq3$ the probability of nucleation of a $S^{D-1}$-topology is found to be more than that of a universe with $S^1\times S^{D-2}$ topology. \\ 
 In 4-dimensions the probabilities reduce to
\begin {equation} 
P_{S^3}\sim e^ {\left[24\pi \alpha +3\pi /\Lambda\right]}  , P_{S^1\times S^2}\sim e^{\left[ 16\pi \alpha +2\pi /\Lambda \right]}.
\end{equation}
A theory with a positive cosmological constant and  $\alpha>0$, leads to a universe without PBH, which is more probable. Though a negative $\alpha<-\frac{1}{8\Lambda }$ leads to greater probability for a universe with PBH, the negative values of $\alpha$ lead to a classical instability in $R^2$-theory [11].\\
(iv) For $\alpha =\gamma =0,\;\beta \neq 0,$ the probabilities  are 
\begin{equation}
P_{S^{D-1}}\sim e^{\left[\frac{2V_{o}I_{D-1}}{H_{o}^D}\left(\frac{-4D(D-1)H_{o}^2+12\Lambda}{D-6}\right)\right]},
\end{equation}
\begin{equation}
P_{S^1\times S^{D-2}}\sim e^{\left[\frac{2V_{o}'(D-3)^{(D-2)/2}}{H^D}\left(\frac{-4DH^2+12\Lambda }{D-6}\right)\right]},
\end{equation}
with $D\neq6$. The probabilities in $D=4$ dimensions may now be obtained as 
\begin{equation}
P_{S^3}\sim e^{\left[\frac{\pi}{H_{o}^4}(H_{o}^2-\Lambda) \right]},P_{S^1\times S^2}\sim e^{\left[\frac{2\pi}{H^4}(4H^2-3\Lambda)\right]}.
\end{equation}
We note that PBH pair creation may be described even without a cosmological constant. The corresponding probabilities are 
\begin{equation}
P_{S^{D-1}}\sim e^{\left[8V_{o}I_{D-1}\left(\frac{\beta}{D-2}\right)^{(D-2)/4}(D(D-1))^{D/2}(6-D)^{(D-6)/4}]\right]},
\end{equation}
\begin{equation}
P_{S^1\times S^{D-2}}\sim e^{\left[8V_{o}'(D-3)^{(D-2)/2}\left(\frac{\beta}{D-2}\right)^{(D-2)/4}D^{D/2}(6-D)^{(D-6)/4}\right]}.
\end{equation}
We note that a universe with  space-time dimensions (i) $D>6$ admits an instanton solution with $\gamma<0$ and (ii) $D<6$,  with a  positive $\beta $. The probability measure for pair creation in inflationary universe is more comparable  to that with a universe without primordial black holes.
\\ Let us consider $D=6$ dimensional universe. In this case $H_{o}^2=\frac{\Lambda }{10}$ in $S^5$-topology and $H^2=\frac{\Lambda }{2}$ in $S^1\times S^4$-topology. The corresponding probabilities are 
\begin{equation}
P_{S^5}\sim e^{\left[\frac{200\pi^2}{3\Lambda^2}\left(1+27\beta \Lambda^2\right)\right]}, P_{S^1\times S^4}\sim e^{\left[\frac{48\pi^2}{5\Lambda^2}\left(1+27\beta \Lambda ^2\right)\right]}.
\end{equation}
de Sitter universe without a pair of PBH is more probable.\\  
(v) For $\Lambda =0,$ the probability measure in the two topologies under consideration are given by
\begin{equation}
P_{S^{D-1}}\sim e^ {\left[\frac{2V_{o}I_{D-1}}{H_{o}^D}\left[\frac{2\alpha D^2(D-1)^2H_{o}^4+4\beta D^3(D-1)^3H_{o}^6+6\gamma D^4(D-1)^4H_{o}^8}{D-2} \right]\right]},
\end{equation}
\begin{equation}
P_{S^1\times S^{D-2}}\sim e^ {\left[\frac{2V_{o}'(D-3)^{(D-2)/2}}{H^D}\left[\frac{2\alpha D^2H^4+4\beta D^3H^6+6\gamma D^4H^8 }{D-2}\right] \right]}.
\end{equation}
In $D=4$ dimensions the probabilities reduce to
\begin{equation}
P_{S^3}\sim e^{\left[\pi(24\alpha +576\beta H_{o}^2+10368\gamma H_{o}^4)\right]},P_{S^1\times S^2}\sim e^{\left[\pi(16\alpha+128\beta H^2+768\gamma H^4)\right]}.
\end{equation}
We now discuss few other special cases which are interesting and follow from the general result obtained above in higher dimensions.\\
(a) For $\gamma=0,$ the probabilities are
\begin{equation}
P_{S^{D-1}}\sim e^ {\left[\frac{2V_{o}I_{D-1}}{H_{o}^D}\left[\frac{2\alpha D^2(D-1)^2H_{o}^4+4\beta D^3(D-1)^3H_{o}^6}{D-2} \right]\right]},
\end{equation}
\begin{equation}
P_{S^1\times S^{D-2}}\sim e^ {\left[\frac{2V_{o}'(D-3)^{(D-2)/2}}{H^D}\left[\frac{2\alpha D^2H^4+4\beta D^3H^6 }{D-2}\right] \right]},
\end{equation}
one determines the instanton solution even with $\Lambda =0$ it reduce to that obtained in {\it Ref. } [9] for $D=4$

(b) For $\beta = 0,$ and $\Lambda =0,$ the probabilities are
\begin{equation}
P_{S^{D-1}}\sim e^ {\left[\frac{2V_{o}I_{D-1}}{H_{o}^D}\left[\frac{2\alpha D^2(D-1)^2H_{o}^4+6\gamma D^4(D-1)^4H_{o}^8}{D-2}\right ]\right]},
\end{equation}
\begin{equation}
P_{S^1\times S^{D-2}}\sim e^ {\left[\frac{2V_{o}'(D-3)^{(D-2)/2}}{H^D}\left[\frac{2\alpha D^2H^4+6\gamma D^4H^8 }{D-2}\right] \right]}.
\end{equation}
For $D=4$
\begin{equation}
P_{S^3}\sim e^{\left[3\pi(8\alpha  +3(128\gamma )^{1/3})\right]},P_{S^1\times S^2}\sim e^{\left[2\pi (8\alpha +3(128\gamma )^{1/3})\right]}.
\end{equation}
Here $H_{o}^2=\sqrt[3]{\frac{1}{3456\gamma }}$ and $H^2=\sqrt[3]{\frac{1}{128\gamma }}$, are new solutions.  In this case a physically interesting instanton solution is obtained for a positive value of $\gamma.$ For $\alpha<0$ and if $|8\alpha |<3(128\gamma )^{1/3},$ de Sitter universe without PBH is more probable. However interesting possibilities emerges when $|8\alpha |>3(128\gamma )^{1/3}$.\\

{\bf 4. Discussions :}\\
In this work we studied the probability of pair creation for primordial black holes in a non linear theory of gravity. We consider an action with a polynomial function of Ricci scalar $R$ as $f(R)=\sum_{i}\lambda_{i}R^i$ to obtain gravitational instanton solutions in the multidimensional universes. The Euclidean action is then evaluated using instanton solutions in the two cases:(i) a universe with $R\times S^d$-topology and (ii) a universe with $R\times S^1\times S^d$-topology in higher dimensions. The former admits an inflationary universe without PBH and later a pair of PBH. We present the corresponding estimation of the probability in a multidimensional universe and compare then to predict the probability of creation of universe. We determine that the probability measures in both the topologies in the framework of a non-linear theory of gravity with square, cubic and quadratic power of scalar curvature term (R)
in the gravitational actions and estimate the coupling parameters for a physically realistic instanton solutions. One interesting point is that in this $f(R)$ gravity model with square and higher order curvature terms one obtains gravitational instanton solution even without a cosmological constant. We note that the de sitter universe without a pair of black holes is more probable for $D\geq3$. We found that the probability of a universe with topology $R\times S^3$- turns out to be much lower than a universe with topology $R\times S^1\times S^2$ in $R^2$-theory unless $\Lambda <-\frac{1}{8\alpha}$ in 4-dimensions. The probability of creation of a universe with a pair of PBH is found to be suppressed if the $R^3$ term is included in the action under the constraints $|\alpha|<2\sqrt{\beta}$ or $\alpha>-2\sqrt{\beta}$ with $\Lambda =0$. Again the probability of creation of a universe with a pair of PBH is found to be  strongly suppressed if one extends the polynomial function of $f(R)$ gravity model upto $R^4$-term with the constraints $|8\alpha |<3(128\gamma )^{1/3}$ or $8\alpha >-3(128\gamma )^{1/3}$.   One obtains Hawking-Turok [15,16] type open inflationary universe in $R\times S^3$-topology, but in  the other topology accommodating PBH does not permit  an open universe instead it gives rise to an anisotropic universe. We note a dimensional dependence of the gravitational instanton and deterimned the different parameters to realize such instantons in the non-linear theory of gravity. A class of new gravitational instantons are presented here which are relevant for cosmological model building.\\

{\bf Acknowledgments :} Authors like to thank
IUCAA Reference Centre at North Bengal University for extending necessary
research facilities to initiate the work. BCP would like to thank UGC, New Delhi for awarding a Project (No. 32-63/2006 (SR) ) and TWAS-UNESCO for awarding visiting Associateship. SG would like to thank University of North Bengal for awarding Junior Research Fellowship.

\end{document}